\documentclass[11pt]{article}

\usepackage[a4paper,margin=1in]{geometry}
\usepackage{lmodern}
\usepackage{braket}
\usepackage{amsmath,amssymb,amsthm}
\usepackage{graphicx}
\usepackage{booktabs}
\usepackage[numbers,sort&compress]{natbib}
\usepackage{doi}
\usepackage{tocloft}
\usepackage{hyperref}

\hypersetup{
  colorlinks=true,
  linkcolor=black,   
  citecolor=blue,
  urlcolor=blue
}

\title{\bfseries
Entanglement Entropy for Screened Interactions via\\
Dimensional Mapping to Harmonic Oscillators
}

\author{
\begin{tabular}{c c}
Akshay Kulkarni\thanks{ \texttt{p20230072@hyderabad.bits-pilani.ac.in}} \, and
\, Rahul Nigam\thanks{ \texttt{rahul.nigam@hyderabad.bits-pilani.ac.in}}
\end{tabular}
\\[0.75ex]
\textit{Birla Institute of Technology and Science Pilani (Hyderabad Campus),}
\\
\textit{Hyderabad 500078, India}
}

\date{}

\date{}

\date{} 

\begin{document}

\maketitle

\begin{center}
{\bf Abstract}
\end{center}
We investigate interaction-induced corrections to entanglement entropy by mapping a
screened Yukawa-type interaction to an effective harmonic oscillator system with
controlled anharmonic perturbations. Starting from a one-dimensional interaction
$V(x) = -g^{2} e^{-\alpha m x}/x$, we reformulate the problem in terms of a
four-dimensional radial oscillator, where the finite screening length generates a
systematic hierarchy of polynomial interactions in the radial coordinate. This mapping
enables a controlled Rayleigh--Schr\"odinger perturbative treatment of the ground-state
wavefunction and an explicit spectral analysis of the reduced density matrix. Working in the weak-screening regime, we compute the leading non-Gaussian correction
arising from the quartic interaction $\rho^{4}$, which appears at order $\alpha^{2}$ in
the expansion of the Yukawa-like potential. We obtain closed analytic expressions for the
resulting small eigenvalues of the reduced density matrix and evaluate their contribution
to the von Neumann entanglement entropy. We show that the entropy receives analytic
corrections at order $\alpha^{2}$, originating both from explicit anharmonic
state-mixing effects and from the implicit $\alpha$-dependence of the Gaussian width
parameter. Our results clarify the distinct roles of harmonic renormalization and genuinely
non-Gaussian interactions in generating entanglement, establish a systematic
power-counting and normalization scheme for higher-order $\rho^{2n}$ perturbations, and
provide a transparent oscillator-based framework for computing entanglement entropy in
weakly interacting low-dimensional and field-theoretic systems.
\newpage
\tableofcontents
\section{\label{sec:Intro}Introduction}
Quantum entanglement in spatially extended systems is both a hallmark of quantum correlations and a powerful diagnostic tool across diverse areas, from quantum information and condensed matter to quantum gravity. Foundational studies \cite{Bombelli1986quantum,Srednicki1993entropy} demonstrated that when the global ground state of a free field is bipartitioned, the reduced von Neumann entropy of one region scales with the area of the boundary, an “area law.” This area-scaling behaviour, first seen in discrete oscillator models and later reinforced in continuum field theory, has become central to our understanding of ground-state structure, tensor-network formulations, and the holographic relation between geometry and entanglement. Entanglement entropy thus serves as a quantitative bridge between microscopic quantum dynamics and macroscopic thermodynamic behavior, offering deep insights into the connections among information, geometry, and energy in quantum systems. In gravitational contexts, it provides a compelling interpretation of the Bekenstein–Hawking entropy as arising from tracing over inaccessible degrees of freedom across horizons. Beyond black holes, entanglement entropy plays a key role in cosmology, many-body systems, and quantum critical phenomena—acting as a diagnostic of correlations, phase transitions, and information flow.

The study of entanglement entropy in discretized field-theoretic models has evolved through several pivotal works. Müller and Lousto \cite{Muller1995entanglement} examined the statistical and geometric aspects of black hole entropy using field discretization techniques, showing that the area law emerges naturally from tracing over field modes inside a horizon. Building on such ideas, Jonker and Vandoren \cite{jonker2016entanglement} developed a Fock-space framework for coupled harmonic oscillators, providing an elegant formalism for computing entanglement entropy beyond the coordinate representation. Their approach clarified the role of mode coupling and provided new insight into the entanglement structure of oscillator networks, thereby strengthening the conceptual link between lattice models and field-theoretic systems. Complementing these advances, the series of works by Das, Shankaranarayanan, Sur, and collaborators established entanglement entropy as a precise probe of the microscopic origin of black hole entropy and its universal features \cite{das2008black,das2008degrees,das2008power}. Through analyses of scalar fields in curved and cosmological backgrounds, they demonstrated that tracing over inaccessible degrees of freedom reproduces the Bekenstein–Hawking area law while revealing universal subleading corrections \cite{das2012entanglement}. Using coupled oscillator models, they constructed a robust computational framework connecting field-theoretic correlations to thermodynamic quantities, showing that entanglement entropy remains invariant under variable redefinitions and is applicable to both static and dynamical spacetimes. Collectively, these studies laid the groundwork for modern explorations of entanglement-based approaches to horizon thermodynamics, quantum gravitational effects, and the emergence of space-time structure from quantum correlations. 

For quadratic (harmonic) systems the reduced density operator is Gaussian and the entanglement spectrum is encoded entirely in the covariance (two-point) data. Powerful analytic and numerical tools exploit this: normal-mode diagonalization, symplectic spectra, and covariance-matrix techniques allow exact calculation of entanglement measures for chains and lattices \cite{Audenaert2002,Weedbrook2012,PeschelEisler2009}. In this work we incorporate weak departures from quadraticity using a perturbative expansion of the relative coordinates Schrödinger equation, following standard asymptotic and perturbative techniques\cite{BenderOrszag1999,ZinnJustin2002}.

These Gaussian techniques underpin much of the intuition about scaling laws, thermalization of subsystems, and transport of correlations in bosonic systems. Realistic systems and many experimentally relevant platforms (cold atoms, trapped ions, nanomechanical resonators, nonlinear optics) contain anharmonicities and nonlinear couplings that drive the state away from Gaussian form. Nonlinearity generically produces nontrivial many-body correlations not captured by covariance information alone. Understanding the leading non-Gaussian corrections to entanglement is crucial for (i) quantifying the limits of Gaussian approximations, (ii) predicting entanglement generation in weakly interacting devices, and (iii) developing analytic control of entanglement scaling in interacting field theories and lattice models. Recent numerical and analytical studies have therefore turned attention to weakly anharmonic chains, finite-temperature behavior, and perturbative approaches to the reduced density matrix. Studies of entanglement in oscillator and many-body systems span a range of analytic techniques and models, the entanglement properties of translationally invariant harmonic chains were characterized using exact covariant methods \cite{Audenaert2002}, foundational scaling behavior of entanglement near quantum phase transitions was explored in spin chains \cite{osborne2002entanglement}, unitary transformations have been used to analyze entanglement in coupled oscillator systems \cite{Jellal_2011}, and long-range harmonic interactions in lattice models have been shown to exhibit rich entanglement structure controlled by temperature and coupling parameters \cite{canovi2014dynamics}. 

In related developments, harmonic and Gaussian systems have continued to provide a controlled setting for exploring entanglement properties in both equilibrium and dynamical situations. Studies of coupled oscillator systems have clarified how entanglement entropy depends on temperature and controlled departures from ground-state configurations, including squeezed excitations \cite{KatsinisPastras2020,KatsinisPastras2023}. Time-dependent protocols, such as global quenches in extended systems, further illustrate how entanglement growth reflects underlying collective structures rather than microscopic details \cite{CalabreseCardy2004}. Complementary approaches based on phase-space descriptions have provided additional intuition for entanglement in systems with low-dimensional effective descriptions \cite{DasHamptonLiu2022}. Related analyses of entanglement scaling and entropy measures highlight how different information measures encode distinct aspects of quantum correlations in such systems \cite{Styliaris2021}.

At the same time, structural aspects of entanglement have been investigated beyond ground states, including rigorous statements about scaling and bounds in oscillator eigenstates \cite{AbdulRahman2024}. Extensions to interacting chaotic systems suggest that many qualitative features of entanglement persist under controlled deformations \cite{Vidmar2017}. Work on operator entanglement has clarified how entanglement measures refine the information contained in quantum states \cite{Zanardi2000}. Developments on symmetry-resolved entropies further demonstrate how additional quantum numbers reorganize entanglement contributions across a wide range of quantum systems \cite{Murciano2021,Calabrese2022,PirmoradianTanhayi2024}.

Closely related analytical techniques for oscillator-based quantum systems have been developed in several complementary directions. Lewis and Riesenfeld provided an exact operator-based formulation for time-dependent quadratic Hamiltonians, establishing invariant-operator methods that produce exact quantum states and spectra for driven oscillators  \cite{lewis1969exact}. Applications to interacting and lattice systems have been pursued in condensed-matter contexts, where entanglement and correlation dynamics of coupled oscillators and bosonic modes were analyzed both numerically and analytically, including weakly anharmonic regimes and quench protocols \cite{song2012bipartite,canovi2014dynamics}. In parallel, field-theoretic studies have employed oscillator and mode-decomposition techniques to extract universal features of entanglement in interacting conformal and non-conformal theories, highlighting the role of perturbative corrections beyond Gaussian order \cite{sivaramakrishnan2018entanglement}. Together, these works underscore the versatility of oscillator-based approaches for accessing entanglement properties in static, dynamical, and weakly interacting quantum systems.

Two complementary perturbative philosophies have been employed in the literature. One is Hamiltonian perturbation theory directly applied to oscillator wavefunctions (Rayleigh–Schrödinger expansions) 
and then using the perturbed wavefunction to form the reduced density matrix, a natural route for low-dimensional systems and discrete chains (examples include specific two-oscillator anharmonic studies). Another is the path-integral/replica-based perturbation approach (or closely related spectral methods) that computes entropy variations by perturbing the action or the geometry and using field-theoretic machinery; Rosenhaus \& Smolkin \cite{rosenhaus2014entanglement} provided a systematic path-integral perturbative framework that has been influential for field-theory settings and in identifying universal terms in entanglement corrections. Both approaches have strengths: the RS route is constructive and explicit for finite systems, while path-integral methods clarify universal terms and connect to QFT techniques. However, most published results either focus on a single perturbation order, remain within Gaussian truncations, or present numerical data without closed prefactors for mixed-order interactions.

Recent developments have also highlighted complementary methodological directions. In particular, invariant-operator techniques (Lewis–Riesenfeld methods) and Ermakov-type approaches have been employed to treat time-dependent and quenched oscillator systems with interactions, yielding analytic, time-resolved reduced density matrices and entanglement spectra (see, e.g., Choudhury et al. \cite{choudhury2022entanglement}). These dynamical studies emphasize that non-Gaussian couplings and quench protocols can produce rich temporal structure in the entanglement spectrum, and they demonstrate that invariant-operator methods provide a powerful complement to static perturbative expansions when one wishes to address driven or time-dependent scenarios.\\
Recent work by Barman and Sardar~\cite{barman2019perturbative} developed a perturbative analysis of entanglement entropy within both Fock and polymer quantization schemes, elucidating how the area law emerges and how it may be modified under alternative quantization prescriptions. Their results underscore both the power and the limitations of perturbative expansions applied to oscillator discretizations of field theories. Complementary numerical studies of anharmonic oscillator chains, including investigations of Rényi entropies, have further explored finite-temperature crossovers and the influence of $\phi^{4}$-type interactions on entanglement diagnostics. Together, these efforts reflect sustained interest in quantifying non-Gaussian entanglement in interacting systems, while also revealing the absence of a systematic algebraic framework capable of handling mixed perturbation orders with explicit analytic control.

Motivated by this gap, and guided by our own explorations, several questions remain incompletely addressed. 
(i) \emph{Mixed-order bookkeeping}: When multiple polynomial perturbations coexist such as $(\alpha^{2}\rho^{4})$, $(\alpha^{3}\rho^{6})$, and $(\alpha^{4}\rho^{8})$, how do lower-order wavefunction amplitudes feed into higher-order corrections to the reduced spectrum, and which combinations dominate the entanglement entropy at a given order? 
(ii) \emph{Analytic prefactors and scaling}: For weak coupling, can one obtain closed-form expressions, and not merely scaling exponents for the leading small eigenvalues of the reduced density matrix, including regular $(\alpha^{r})$ term contributions to the entropy?
(iii) \emph{Gaussian versus non-Gaussian regimes}: Under what conditions does the covariance (Gaussian) approximation suffice for entanglement measures, and when are genuinely non-Gaussian corrections indispensable? 
(iv) \emph{Extensibility}: How do these structures generalize to systems with more degrees of freedom, such as short chains, normal-mode decompositions, and finite-temperature states?

These questions are of direct relevance to both theoretical developments, ranging from perturbative quantum field theory to corrections to area laws—and experimental platforms where weak nonlinearities are unavoidable, including trapped ions, superconducting circuits, and nanomechanical systems. While previous studies have addressed isolated aspects of these issues through order-by-order perturbative recipes, numerical chain calculations, or replica-based field-theoretic approaches, a compact, modular, and reusable algebraic scheme that yields explicit coefficients for mixed perturbative orders has remained elusive.

In this work, we present such a framework. We (i) derive a Rayleigh–Schrödinger recursion tailored to Hamiltonians containing multiple polynomial perturbations of the form $\sum_{r}(-1)^{r}\alpha^{r}V^{(r)}$; (ii) compute ground-state amplitude corrections explicitly up to $\mathcal{O}(\alpha^{4})$, carefully accounting for cross-order contributions (e.g., first-order effects of $V^{(4)}$ and second-order insertions of $V^{(2)}$); (iii) construct the full density operator and perform the partial trace to obtain the reduced operator in a compact block form, with nontrivial structure encoded in a rectangular amplitude matrix $\mathcal{C}$; (iv) demonstrate analytically how the singular values of $\mathcal{C}$ control the small eigenvalues of the reduced density matrix and yield closed-form prefactors for the von Neumann entropy expansion; (v) benchmark these results against Gaussian (covariance-based) predictions to isolate genuinely non-Gaussian effects; and (vi) validate the analytic expressions using truncated-basis numerics, while outlining extensions to short chains and finite-temperature settings. 

Our approach combines the concreteness of wavefunction-based perturbation theory with the spectral clarity of singular-value analysis, thereby extending earlier perturbative studies \cite{barman2019perturbative, rosenhaus2014entanglement}
and providing a systematic route to entanglement entropy beyond the Gaussian regime. From a quantum-information perspective, the present framework provides an analytic handle on interaction-induced entanglement in few-body continuous-variable systems. In platforms such as trapped ions and cold atoms confined in optical potentials, effective interparticle interactions are often of finite range and can be well approximated by Yukawa-type forms. Expanding these interactions about equilibrium naturally generates anharmonic corrections to an otherwise harmonic relative mode. While Gaussian analyses based on covariance matrices capture the leading harmonic entanglement, they are insensitive to the genuinely non-Gaussian correlations induced by such anharmonicities. Our results demonstrate explicitly how quartic terms in the effective interaction generate new, parametrically small eigenvalues of the reduced density matrix and produce non-analytic corrections to the entanglement entropy. This provides a systematic way to quantify when Gaussian approximations break down and offers a route to using entanglement entropy as a diagnostic of finite-range interaction effects in few-body quantum systems.

Following the mapping of one-dimensional Coulomb problem to a four-dimensional radial oscillator introduced by Bateman et al. \cite{bateman1992} and subsequently implemented by S. Shankaranarayanan et al. \cite{Shankydivergence} for the hydrogen atom, we employ a related mapping framework to treat Yukawa-like interactions via a similar four-dimensional oscillator construction.

The paper is organized as follows. In Sec. II we map the screened Yukawa interaction to an effective four-dimensional harmonic oscillator with anharmonic perturbations. In Sections III \& IV we compute the perturbed ground state and reduced density matrix explicitly at quartic order. Section V develops a systematic perturbative expansion of the entanglement entropy, clearly separating Gaussian and non-Gaussian contributions. In Sec. VI we generalize the analysis to arbitrary $\rho^{2n}$ perturbations and establish universal power-counting rules. We conclude with implications and outlook.

\section{\label{sec:MapOsc}Map to coupled oscillator}
We begin by considering a one-dimensional form of the potential,

\begin{equation}
V(x) = -g^2 \frac{e^{-\alpha m x}}{x},
\end{equation}
where g denotes the coupling strength, $\alpha$ is a dimensionless screening parameter, and $m$ represents the mass of the mediating particle that governs the interaction range. Although the true Yukawa potential arises as the Green’s function of a massive scalar field in three dimensions, we consider here a one-dimensional "Yukawa-like" screened potential. This choice preserves the key qualitative feature of screening through the exponential decay term while retaining the $1/x$ dependence, ensuring continuity with the long-range limit.
The Yukawa-like potential serves as a paradigmatic model for screened interactions across a broad range of physical systems—from nuclear and plasma physics to condensed-matter and quantum field settings. The exponential suppression term $e^{-\alpha m x}$ introduces a natural length scale $(\alpha m)^{-1}$, which allows a smooth interpolation between two physically distinct regimes: in the limit $\alpha \to 0$, the potential reduces to the long-range $1/x$ interaction, whereas for finite $\alpha$, it captures short-range screened behavior characteristic of massive field mediators.

In the present study, we restrict attention to the weak-screening regime $\alpha \ll 1$, treating the screening parameter as a small perturbative quantity. Our objective is to determine how such screening effects modify the entanglement entropy, expanding the corrections systematically as a power series in $\alpha$. Instead of approaching this problem directly through field-theoretic or replica-based formulations, which often involve technically demanding functional determinants or analytic continuations, we adopt an alternative route: mapping the potential to an equivalent coupled harmonic oscillator system.

This oscillator mapping offers several conceptual and computational benefits. First, it provides a clear physical picture of entanglement as mode coupling between discrete degrees of freedom. In this representation, the interaction potential translates into explicit couplings among oscillators, encapsulated in a symmetric coupling matrix whose eigenvalues directly determine the entanglement spectrum. The task of computing the reduced density matrix and the von Neumann entropy thereby reduces to a linear-algebraic problem involving covariance or coupling matrices, rendering the procedure both transparent and analytically tractable.

Furthermore, perturbative corrections arising from the potential form can be incorporated systematically as polynomial modifications to the oscillator coupling matrix, enabling a controlled expansion in powers of $\alpha$ or g. This framework thus permits a uniform treatment of leading and higher-order corrections to entanglement measures, with results that can be verified numerically using truncated oscillator bases. Conceptually, the oscillator mapping also connects naturally with gravitational analogues: tracing over one subset of oscillators across a spatial partition parallels tracing over inaccessible field modes across a causal horizon, thereby linking our analysis to the well-established entanglement interpretation of black hole entropy.

In essence, the harmonic-oscillator mapping furnishes a unifying and physically transparent framework that combines analytic control, numerical flexibility, and conceptual clarity, making it particularly well-suited for probing entanglement entropy in systems governed by screened or weakly interacting potentials such as the Yukawa form.
The Schrodinger equation for a two particle system in given potential is

\begin{eqnarray} \label{Sch1d}
\frac{- \hbar^2}{2m_1} \frac{\partial^2 \Psi(x_1,x_2)}{\partial x_1^2 } - \frac{ \hbar^2}{2m_2}\frac{\partial^2 \Psi(x_1,x_2)}{\partial x_2^2}  -g^2\frac{e^{-\alpha m |x_1 - x_2|}}{|x_1 - x_2|} \Psi(x_1,x_2) = -B \Psi(x_1,x_2)
\end{eqnarray}
where B is the absolute value of binding energy, $m_i$ is mass of the interacting particles, g is the coupling constant and $\alpha$ is the screening paramater.

We switch to the centre of mass and relative coordinates $x_c$ and $x$ respectively.
\begin{equation}
x_c = \frac{m_1 x_1 + m_2 x_2}{m_1 + m_2}, \quad x = x_1 - x_2
\end{equation}
This makes
\begin{align}
x_1 = \frac{M x_c + m_2 x}{M}, \quad
x_2 = \frac{M x_c - m_1 x}{M}
\end{align}
where $M = m_1 + m_2$.
The schrodinger equation now becomes,
\begin{equation}\label{relScheq}
-\frac{\hbar^2}{2M}\frac{\partial^2 \Psi}{\partial x_c^2} - \frac{\hbar^2}{2\mu}\frac{\partial^2 \Psi}{\partial x^2} - g^2 \frac{e^{-\alpha m |x|}}{|x|} \Psi = - B \Psi
\end{equation}
where $\mu = m_1 m_2 /(m_1 + m_2)$ is the reduced mass of the system.
Using separation of variables method, let
\begin{equation}
\Psi(x,x_c) = f(x)g(x_c)
\end{equation}
we obtain,
\begin{eqnarray}
-\frac{\hbar^2}{2M}\frac{1}{g(x_c)}\frac{\partial^2 g(x_c)}{\partial x_c^2} -\frac{\hbar^2}{2\mu}\frac{1}{f(x)}\frac{\partial^2 f(x)}{\partial x_c^2}
- g^2 \frac{e^{-\alpha m |x|}}{|x|} = -B
\end{eqnarray}
This results in two equations, one for centre of mass and other in terms of relative coordinates as
\begin{equation}
-\frac{\hbar^2}{2M}\frac{\partial^2 g(x_c)}{\partial x_c^2} = E_c\, g(x_c)
\end{equation}
and
\begin{equation}\label{Sch2particle}
-\frac{\hbar^2}{2\mu}\frac{\partial^2 f(x)}{\partial x^2} - g^2 \frac{e^{-\alpha m |x|}}{|x|} f(x) = -B_r\, f(x) 
\end{equation}
after separation, the total energy is written as $E = E_c - B_r$, where $E_c$ being the center of mass kinetic energy and $B_r >0$ is the binding energy associated with the relative motion. While the first equation describes free particle motion, our interest lies in the second equation which can be used for describing both the bound as well as scattering states. Since $x$ and $x_c$ both are now independent coordinates, we can freely use non-linear transformation without the loss of generality and restore the original coordinates at the end of derivation to obtain the entangled wavefunction and density operator.

In order to transform the above equation to the form of harmonic oscillator, we use
\begin{equation}
x = \rho^2
\end{equation}
using this transformation, eq.(\ref{Sch2particle}) becomes
\begin{equation}
\label{Sch1da}
-\frac{\hbar^2}{2\mu} \left(\frac{d^2f}{d\rho^2} - \frac{1}{\rho}\frac{df}{d\rho} \right) + 4B_r\rho^2 f = 4g^2 e^{-\alpha m \rho^2}f
\end{equation}
This is yet not an eigenvalue equation representing the harmonic oscillator due to presence of first order derivative in addition to the exponential factor. We seek to find a transformation of $f(\rho)$ that would map eq.(\ref{Sch1da}) to a d-dimensional harmonic oscillator of the form - 
\begin{eqnarray}\label{Hosc3d}
-\frac{\hbar^2}{2\mu} \left(\frac{d^2}{dr^2} - \frac{(d-1)(d-3)}{4r^2} - \frac{l(l+1)}{r^2}\right)u(r) 
+ \frac{1}{2}\mu\omega^2 r^2 u(r) = E u(r)
\end{eqnarray}

Further defining(See Appendix A.2 for detailed derivation)
\begin{equation}\label{wavefnfinal}
f(\rho) = A \sqrt{\rho} \phi(\rho)
\end{equation}
and taking its first and second order derivatives -  
\begin{equation}\label{Psi'}
f'(\rho) = \frac{A}{2\sqrt{\rho}} + A\sqrt{\rho} \phi'(\rho)
\end{equation}
\begin{align}\label{Psi''}
f''(\rho) = &\frac{-A \phi}{4 \rho^{3/2}}   
+ \frac{A\phi'}{\sqrt{\rho}} + A\sqrt{\rho}\phi''
\end{align}
 Substituting eqs.(\ref{wavefnfinal}),(\ref{Psi'}) and (\ref{Psi''}) in eq.(\ref{Sch1da}), we get
 \begin{align}
 \frac{-\hbar^2}{2\mu} \left( -\frac{\phi}{4\rho^{3/2}} + \sqrt{\rho}\phi'' - \frac{\phi}{2\rho^{3/2}} \right) + 4B_r\rho^2 \sqrt{\rho} \phi = 4g^2 e^{-\alpha m \rho^2} \sqrt{\rho}\phi
\end{align}
finally leading to
\begin{eqnarray}
\nonumber\frac{-\hbar^2}{2\mu} \left( \frac{d^2\phi}{d\rho^2} - \frac{3}{4} \frac{\phi}{\rho^2} \right) + \, 4B_r\rho^2\phi 
= 4g^2 e^{-\alpha m\rho^2} \phi
\end{eqnarray}
This equation is similar to 4-d radial harmonic oscillator equation eq.(\ref{Hosc3d}) for $l = 0$, except for the exponential term which also prevents it from being an eigenvalue equation. This can be resolved by considering perturbations to the 4-d harmonic oscillator by a series of anharmonic terms. We want to find the perturbed ground state wavefunction which would allow us to compute the density operator and then the entanglement entropy. In the rest of this work, we will study the effects of anharmonic perturbations to the entanglement entropy. A typical perturbation term in the series is of the form $\alpha^n m^n \rho^{2n}/n!$.  \\

\section{Perturbed Ground State}

Before we proceed to study the effects of generic terms, we first illustrate the technique by expanding up to $\alpha^2 m^2 \rho^{4}/2!$ term. We get
\begin{eqnarray}
\frac{-\hbar^2}{2\mu} \left( \frac{d^2\phi}{d\rho^2} - \frac{3}{4} \frac{\phi}{\rho^2} \right) + 4B_r\rho^2\phi \, = 4g^2\phi -4g^2 \alpha
 m \rho^2 \phi  + \frac{1}{2!}4g^2 \alpha^2 m^2 \rho^4 \phi
\end{eqnarray}
The term linear in $\alpha$ is quadratic in the coordinate $\rho$ and therefore preserves the harmonic nature of the Hamiltonian. Its only effect is to renormalize the oscillator frequency (or equivalently the coefficient of $\rho^2$), shifting the zeroth–order spectrum and the Gaussian width of the ground–state wavefunction. Since quadratic Hamiltonians remain exactly solvable, this contribution does not introduce genuinely new non-Gaussian correlations, rather it modifies the covariance structure of the ground state. For this reason, and to clearly isolate interaction–induced (non-Gaussian) effects, it is convenient to absorb the $\mathcal O(\alpha)$ harmonic term into a redefined unperturbed Hamiltonian. Rewriting the equation accordingly, we obtain - 
\begin{eqnarray}
\frac{-\hbar^2}{2\mu} \left( \frac{d^2\phi}{d\rho^2} - \frac{3}{4} \frac{\phi}{\rho^2} \right) + 4B_r'\rho^2 \phi - 2g^2 \alpha^2 m^2 \rho^4 \phi
= 4g^2\phi  
\end{eqnarray}
where the renormalized harmonic coefficient is $B_r' = (B_r +g^2 \alpha
 m)$. The remaining perturbation is therefore purely anharmonic -
 \begin{equation}
 H' = 2g^2 \alpha^2 m^2 \rho^4
 \end{equation}
At this point it is important to emphasize a subtle but conceptually significant distinction. If the quadratic $\alpha\rho^2$ term were instead treated perturbatively, its second–order contribution to the ground–state wavefunction would generate corrections of order $\alpha^2$. These corrections, however, are entirely Gaussian in nature and precisely reproduce the $\alpha^2$ terms obtained by expanding observables (including entanglement entropy) in the renormalized frequency $B_r'$. Thus, whether the $\alpha\rho^2$ term is handled exactly or perturbatively, its effect on the entanglement entropy is fully captured by Gaussian covariance data and does not produce new eigenvalues in the reduced density matrix. By contrast, the $\rho^4$ term represents a genuinely non-Gaussian perturbation. Even at first order in perturbation theory, it induces admixtures of excited oscillator states into the ground state, leading to new small eigenvalues in the reduced density matrix and to characteristic non-analytic contributions to the entanglement entropy.
Separating the Gaussian frequency renormalization from the anharmonic interaction in this manner avoids double counting and makes transparent which contributions to the entropy arise from simple covariance renormalization and which reflect intrinsically non-Gaussian correlations. This reorganization of the perturbation theory provides a clean and systematic framework for the analysis that follows and naturally generalizes to higher–order polynomial interactions, where multiple Gaussian and non-Gaussian contributions coexist and must be carefully disentangled.\\
We begin by defining the effective oscillator frequency and Gaussian width parameter as
\begin{equation}
\omega_0 \equiv \sqrt{\frac{8 B_r'}{\mu}}, 
\quad 
\beta \equiv \frac{\mu \omega_0}{\hbar}.
\end{equation}
With these definitions, the unperturbed radial Hamiltonian corresponds to a four-dimensional isotropic harmonic oscillator. The normalized ground-state radial wavefunction is
\begin{equation}
\phi_0(\rho) = \left(\frac{\beta}{\pi}\right) e^{-\frac{1}{2}\beta \rho^2},
\end{equation}
with unperturbed energy eigenvalue
\begin{equation}
E_0^{(0)} = 2\hbar \omega_0 .
\end{equation}
The complete set of radial eigenfunctions in four dimensions is given by
\begin{equation}
\phi_{n_r,\ell}(\rho) =
\sqrt{\frac{2\, \beta^{\ell+2} \, n_r!}{\Gamma(n_r + \ell + 2)}} \;
\rho^\ell \, e^{-\tfrac{1}{2}\beta \rho^2} \,
L_{n_r}^{(\ell+1)}(\beta \rho^2),
\end{equation}
where $n_r$ is the radial quantum number, $\ell$ is the angular momentum, and
$L_{n_r}^{(\ell+1)}$ are associated Laguerre polynomials.  
Since the mapping leads to a spherically symmetric effective problem, we restrict attention to the $\ell=0$ sector throughout.
The first-order correction to the $n$th eigenstate due to a perturbation $H'$ is given by standard Rayleigh–Schrödinger perturbation theory,
\begin{equation}
\phi_n^{(1)}(\rho)
=
\sum_{n_r\neq n}
\frac{\langle \phi_{n_r}^{(0)} | H' | \phi_n^{(0)} \rangle}
{E_n^{(0)} - E_{n_r}^{(0)}} \,
\phi_j^{(0)}(\rho).
\label{Pert}
\end{equation}
In the present case, the perturbation arises from the quartic term generated by the Yukawa expansion,
\begin{equation}
H' = 2 g^2 \alpha^2 m^2 \rho^4 .
\end{equation}
Because $\rho^4$ is a scalar operator, it preserves angular momentum and couples only states with the same $\ell$. Consequently, only $\ell=0$ radial excitations contribute to the correction of the ground state.
Introducing the dimensionless variable $x = \beta \rho^2$, the relevant matrix elements reduce to
\begin{equation}
\langle n_r | \rho^4 | 0 \rangle
=
2\pi^2 \, \frac{N_{n_r} N_0}{2\beta^4}\, I_{n_r},
\end{equation}
where
\begin{equation}
I_{n_r} = \int_0^\infty e^{-x} x^3 L_{n_r}^{(1)}(x)\, dx .
\end{equation}
Evaluating these integrals yields
\begin{equation}
I_0 = 6, \qquad
I_1 = -12, \qquad
I_2 = 6, \qquad
I_3 = 0 ,
\end{equation}
implying that only the $n_r = 1,2$ states contribute at this order.
The corresponding energy denominators are
\begin{equation}
E_0^{(0)} - E_1^{(0)} = -2\hbar \omega_0, 
\qquad
E_0^{(0)} - E_2^{(0)} = -4\hbar \omega_0 .
\end{equation}
Combining these results, the first-order expansion coefficients are
\begin{align}
a_1 = -\frac{12\sqrt{2}\, g^2 \pi^2 \alpha^2 m^2}
{\beta^2 \hbar \omega_0}, \quad
a_2 = \frac{2\sqrt{3}\, g^2 \pi^2 \alpha^2 m^2}
{\beta^2 \hbar \omega_0}.
\end{align}
Using $\beta = \mu \omega_0 / \hbar$, these may be equivalently written as
\begin{align}
a_1 = -12\sqrt{2}\,
\frac{\hbar g^2 \pi^2 \alpha^2 m^2}
{\mu^2 \omega_0^3},\quad
a_2 = 2\sqrt{3}\,
\frac{\hbar g^2 \pi^2 \alpha^2 m^2}
{\mu^2 \omega_0^3}.
\end{align}
The corrected ground-state wavefunction, accurate to $\mathcal{O}(\alpha^2)$, therefore takes the form
\begin{equation}
\phi_0^{(1)}(\rho)
\simeq
\phi_0(\rho) + a_1 \phi_1(\rho) + a_2 \phi_2(\rho).
\end{equation}
Upon combining terms, this can be written compactly as
\begin{equation}
\phi_0^{(1)}(\rho)
=
\mathcal{N}\,
e^{-\frac{1}{2}\beta \rho^2}
\left(
c_0 + c_2 \rho^2 + c_4 \rho^4
\right),
\end{equation}
where the coefficients are
\begin{align}
c_0 = \sqrt{2}\,\beta
- 18 \sqrt{2}\,
\frac{g^2 \pi^2 \alpha^2 m^2}
{\hbar \omega_0 \beta}, \,
c_2 = 6 \sqrt{2}\, 
\frac{g^2 \pi^2 \alpha^2 m^2}
{\hbar \omega_0}, \,
c_4 = \sqrt{2}\,\beta\,
\frac{g^2 \pi^2 \alpha^2 m^2}
{\hbar \omega_0}.
\end{align}
This explicit polynomial–Gaussian structure of the corrected wavefunction will be central to the construction of the reduced density matrix and the subsequent computation of entanglement entropy.
So the corrected wavefunction in terms of radial coordinates $r_1$ and $r_2$ is given by
\begin{align}
\phi_0^1(r_1,r_2) &=\,
e^{-\frac{1}{2}\beta (r_1 - r_2)^2}
\Biggl[
\sqrt{2}\beta
- 18\sqrt{2}\,\frac{g^2 \pi^2 \alpha^2 m^2}{\hbar \omega_0 \beta}
 + 6\sqrt{2}\frac{g^2 \pi^2 \alpha^2 m^2}{\hbar \omega_0}(r_1 - r_2)^2 \notag \\
&\quad
+ 2\sqrt{2}\beta \frac{g^2 \pi^2 \alpha^2 m^2}{2 \hbar \omega_0} (r_1 - r_2)^4
\Biggr]
\end{align}

At this stage, the original two-body problem with a Yukawa-like interaction has been mapped onto an effective radial harmonic oscillator in four dimensions, perturbed by anharmonic terms originating from the screened interaction. This mapping is exact for the relative-coordinate Schrodinger equation in the s-wave sector and provides a faithful reformulation of the problem for the purposes of computing ground state entanglement. The resulting oscillator framework provides a controlled and analytically tractable setting in which both the ground-state wavefunction and its perturbative corrections can be computed systematically. Importantly, this reformulation allows the entanglement between the two particles to be analyzed directly through the structure of the resulting wavefunction, paving the way for an explicit construction of the reduced density matrix and the associated entanglement entropy.

\section{Reduced Density Matrix}

Having established the mapping to an effective four-dimensional harmonic oscillator, we now turn to the computation of the reduced density matrix associated with the ground state of the system. For the unperturbed oscillator, the ground-state wavefunction is Gaussian in the relative coordinate, implying that the reduced density matrix inherits a translationally invariant Gaussian structure. This property allows its eigenvalues to be obtained analytically through a Fourier transformation, thereby providing a closed-form expression for the entanglement entropy. In the subsequent analysis, we first review this Gaussian baseline and then incorporate perturbative corrections arising from anharmonic terms in a systematic expansion.
The reduced density operator is given as: 
\begin{align}
\nonumber\rho(r_1,r'_1) &= \int_0^{\infty} \phi_0^1(r_1,r_2)\phi_0^{*1}(r'_1,r_2) d^4 r_2 
\end{align}
Before we compute this integral, we need to define our setup. Consider $r'_1$ and $r_1$ to be at $(-s/2,0,0,0)$ and $(+s/2,0,0,0)$ and let $r_2$ be at distance y from the origin making an angle (polar angle) $\theta$ with the line joining $r_1$ and $r'_1$. So we have 
\begin{eqnarray}
|r_1 - r_2| = r_- = \sqrt{y^2 + s^2/4 - y s cos \theta} \\ \nonumber
|r'_1 - r_2| = r_+ = \sqrt{y^2 + s^2/4 + y s cos \theta}
\end{eqnarray}
And the measure $d^4 r_2 = 2 \pi^2 y^3 dy sin^2 \theta d \theta$. In this notation, the reduced density operator can be written as 
\begin{align}
\rho(s) &= 2\pi^2\int_0^{\infty} \phi_0^1(y,s,\theta)\phi_0^{*1}(y,s,\theta) \nonumber \, sin^2\theta \, y^3\,dy \,d\theta \\ 
&= 2 \pi^2 e^{-\beta s^2/4} \int_0^{\infty} y^3 dy \int_0^{\pi} sin^2 \theta d\theta e^{-\beta y^2} P(r_+) P(r_-) 
\end{align}
Here the product $P(r_+)P(r_-)$ is of the form
\begin{equation}
\bigg[c_o + c_2 r_+^2 + c_4 r_+^4\bigg]\bigg[c_o + c_2 r_-^2 + c_4 r_-^4\bigg]
\end{equation}
As we restrict to the order $\mathcal{O}(\alpha^2)$, the relevant terms in this product would be 
\begin{equation}
c_0^2 + c_0c_2 (r_+^2 + r_-^2) + c_0c_4 (r_+^4 + r_-^4)
\end{equation}

First we carry out the $\theta$ integral to get 
\begin{align}
\rho(s) = 2 \pi^2 e^{-\beta s^2/4} \int_0^{\infty} y^3 dy e^{-\beta y^2} &\bigg[ \pi c_0^2/2 + 
\pi c_0c_2 (y^2 + s^2/4) + \pi c_0 c_4 [(y^2 + s^2/4)^2 \\&+ y^2 s^2/4 ] \bigg] \nonumber
\end{align}
and then integrating over y, we get
\begin{eqnarray}
\rho(s) = 
c_0 \pi^3 e^{-\beta s^2/4}  \frac{\left( 96 c_4 + 
   \beta \left( 8 \beta c_0 + 32 c_2 + 4 (\beta c_2 + 6 c_4) s^2 + \beta c_4 s^4 \right) \right)}
{16 \beta^4}
\end{eqnarray}
Writing the density operator explicitly and restricting it up to $\mathcal{O}(\alpha^2)$, we get
\begin{align}
\rho(s) = \pi^3 e^{-\beta s^2 /4} + \frac{e^{-\beta s^2/4} g^2 m^2 \pi^2 \alpha^2 s^2(48 + \beta s^2)}{8 \beta \hbar \omega_0}
\end{align}
By using fourier transform, we can obtain the eigenvalues of the reduced density operator $\rho(s)$ and since s is a continuous variable, the eigenvalue will be continuous in some variable k. We demonstrate the procedure below,
\begin{align}
\int_{-\infty}^{+\infty} \rho(s) e^{-iks} ds = \tilde{\rho}(k)
\end{align}
which essentially translates as 
\begin{align}\label{Rk}
\int_{-\infty}^{+\infty} \rho(r-r')e^{-ikr'} dr' = \tilde{\rho}(k)e^{-ikr}
\end{align}
resulting in
\begin{align}
\tilde{\rho}(k) =
\frac{e^{-k^2/\beta} \pi^{7/2}}{\beta^{9/2} \hbar \omega_0}
\Big(2 \,\beta^{4}\, \hbar \, \omega_0
    + 2\, g^{2} m^{2} \pi^{2} \alpha^{2}
    \big[
        27\beta^2 - 60 \beta k^2 + 4k^4
    \big]
\Big)
\end{align}
Since the reduced density matrix is translationally invariant and has a continuous spectrum, we will define the entanglement entropy later by using normalized momentum-space eigenvalue given by
\begin{align} \label{rhonorm}
\rho(k) =
\frac{e^{-k^2/\beta}}{2\beta^{9/2} \pi^{1/2} \hbar \omega_0}
\Big(2 \,\beta^{4}\, \hbar \, \omega_0
    + 2\, g^{2} m^{2} \pi^{2} \alpha^{2}\big[
        27\beta^2 - 60 \beta k^2 + 4k^4
    \big]
\Big)
\end{align}

At this stage, it is crucial to stress that the expression for the reduced density matrix eigenvalue obtained above does not yet constitute an explicit perturbative expansion in the screening parameter $\alpha$, despite the appearance of $\alpha^{2}$ in the prefactor. The reason is that the Gaussian width parameter $\beta$, which controls both the exponential decay and the normalization of $\tilde{\rho}(k)$, itself carries implicit $\alpha$-dependence through the renormalized harmonic coefficient
\begin{equation}
B_r' \;=\; B_r + g^{2}\alpha \,m  .
\end{equation}
Since $\beta$ is determined by $B_r'$ via the unperturbed oscillator spectrum, we must regard it as a function $\beta(\alpha)$ rather than a constant.\\
To make this dependence explicit, we expand $\beta(\alpha)$ perturbatively around its unperturbed value $\beta_0 \equiv \beta(\alpha=0)$,
\begin{equation} \label{betaexp}
\beta(\alpha)
= \beta_0 + \alpha\,\beta_1 + \alpha^{2}\beta_2 + \mathcal{O}(\alpha^{3}) \, ,
\end{equation}
with
\begin{align}
\beta_0 = \frac{\mu}{\hbar}\sqrt{\frac{8B_r}{\mu}}, \quad
\beta_1 = \frac{g^2 m}{2B_r}\,\beta_0, \quad
\beta_2 = -\frac{1}{8}\left(\frac{g^2 m}{B_r}\right)^2 \beta_0 .
\end{align}
The coefficients $\beta_1$ and $\beta_2$ encode purely Gaussian renormalization effects arising from the harmonic ($\rho^2$) deformation.
Substituting this expansion into the eigenvalue $\rho(k)$, all $\beta$-dependent structures must be expanded consistently. For example, the Gaussian kernel becomes
\begin{align}
e^{-k^{2}/\beta(\alpha)}
&=
e^{-k^{2}/\beta_0}
\left[
1 + \alpha\,\frac{\beta_1 k^{2}}{\beta_0^{2}}
+ \alpha^{2}\!\left(
\frac{\beta_2 k^{2}}{\beta_0^{2}}
- \frac{\beta_1^{2} k^{2}}{\beta_0^{3}}
  + \frac{\beta_1^{2} k^{4}}{2\beta_0^{4}}
\right)
+ \mathcal{O}(\alpha^{3})
\right]
\end{align}
and analogous expansions apply to the normalization factors involving powers of $\beta$.

This procedure reveals that contributions of order $\alpha^{2}$ arise from two distinct sources: (i) explicit anharmonic terms proportional to $\rho^{4}$ in the Hamiltonian, and (ii) implicit corrections generated by the $\alpha$-dependence of the Gaussian width through the harmonic frequency renormalization. While the latter preserve the Gaussian structure of the ground state and may be interpreted as frequency renormalization effects, the former encode genuinely non-Gaussian correlations and lead to qualitatively new contributions to the entanglement spectrum.

Only after all $\beta$-dependent terms have been expanded and collected can the eigenvalue be written in the schematic form
\begin{equation} \label{rhoexp}
\rho(k)
=
\rho^{(0)}(k)
+ \alpha\,\rho^{(1)}(k)
+ \alpha^{2}\,\rho^{(2)}(k)
+ \mathcal{O}(\alpha^{3}) ,
\end{equation} 
which then permits a controlled expansion of the entanglement entropy. This separation is essential for correctly identifying the relative contributions of Gaussian frequency renormalization and non-Gaussian interactions to the entanglement entropy

It is important to note the consequences of the normalization of the momentum space eigenvalue.
By construction, the reduced eigenvalue $\rho(k)$ is normalized to unity,
\begin{equation}
\int_{-\infty}^{\infty} dk \, \rho(k) = 1 ,
\label{eq:rho_norm_full}
\end{equation}
order by order in the perturbative expansion in the screening parameter $\alpha$.
Using the expansion in eq.\eqref{rhoexp}
and inserting it into eq.~\eqref{eq:rho_norm_full}, one obtains the hierarchy of normalization conditions
\begin{align}\label{eq:rho_norm} 
\int dk \, \rho^{(0)}(k) = 1, \quad
\int dk \, \rho^{(1)}(k) = 0, \quad
\int dk \, \rho^{(2)}(k) = 0, 
\end{align}
and similarly for all higher orders.
The zeroth-order condition in eq.~\eqref{eq:rho_norm} fixes the normalization of the unperturbed Gaussian eigenvalue, while the higher-order constraints in eq.\eqref{eq:rho_norm} ensure that perturbative corrections only redistribute spectral weight without changing the total probability.

These relations play a crucial role in the entropy calculation.
In particular, they guarantee that linear and quadratic corrections to the von Neumann entropy may be consistently simplified using integration by parts and Gaussian moment identities, and they ensure that spurious constant contributions do not appear in $S^{(1)}$ and $S^{(2)}$.
Throughout the remainder of this work, we explicitly enforce the normalization conditions in eq.\eqref{eq:rho_norm}, which serve as nontrivial consistency checks on the perturbative density matrix and its Gaussian and non-Gaussian components.

As in eq.\eqref{rhonorm}, the eigenvalue of the reduced density matrix obtained from the partially traced ground state takes the form
\begin{align}
\rho(k)
&=
\frac{e^{-k^{2}/\beta}}{2\, \beta^{9/2} \,\pi^{1/2} \, \hbar \,\omega_0}
\Bigg[
2 \beta^4 \hbar \omega_0
+ g^{2} m^{2} \pi^{2} \alpha^{2}
\mathcal{P}(\beta,k)
\Bigg]
\label{eq:rho_full}
\end{align}
where
\begin{equation}
\mathcal{P}(\beta,k)
= 27\beta^2 - 60\beta k^2 + 4k^4
\end{equation}
Collecting terms of order $\alpha^{0}$ and using eq.\eqref{eq:rho_full}, we obtain the unperturbed eigenvalue
\begin{equation}
\rho^{(0)}(k)
=
\frac{\exp\!\left(-\frac{k^{2}}{\beta_{0}}\right)}{\sqrt{\pi\beta_0}}
\label{eq:rho0}
\end{equation}
This is the purely Gaussian contribution arising from the harmonic oscillator ground state.

At order $\alpha$, only the $\alpha$-dependence of $\beta$ contributes. Expanding the exponential and normalization factors to linear order yields
\begin{equation}
\rho^{(1)}(k)
=
\rho^{(0)}(k)
\left[
\frac{\beta_{1}}{\beta_{0}^{2}}\,k^{2}
-
\frac{1}{2}\frac{\beta_{1}}{\beta_{0}}
\right]
\label{eq:rho1}
\end{equation}
This correction originates entirely from the harmonic ($\rho^{2}$) perturbation and corresponds to a renormalization of the Gaussian width. It does not introduce genuinely non-Gaussian correlations. At order $\alpha^{2}$, two distinct contributions arise.\\
(i) Gaussian renormalization.
Expanding the Gaussian factors to second order gives
\begin{equation}
\rho^{(2)}_{\mathrm{G}}(k)
=
\rho^{(0)}(k)
\Big[
A_{0}
+
A_{2} k^{2}
+
A_{4} k^{4}
\Big],
\label{eq:rho2G}
\end{equation}
with coefficients
\begin{align}
A_{0} =
\frac{3}{8}\frac{\beta_{1}^{2}}{\beta_{0}^{2}}
-
\frac{1}{2}\frac{\beta_{2}}{\beta_{0}}, \quad
A_{2} =
\frac{\beta_{2}}{\beta_{0}^{2}}
-
\frac{3}{2}\frac{\beta_{1}^{2}}{\beta_{0}^{3}}, \quad
A_{4} =
\frac{1}{2}\frac{\beta_{1}^{2}}{\beta_{0}^{4}}
\end{align}
These terms reflect higher-order renormalization of the Gaussian structure and may still be absorbed into an effective width at this order.\\
(ii) Genuine non-Gaussian contribution. The explicit anharmonic $\rho^{4}$ perturbation contributes at order $\alpha^{2}$ as
\begin{equation}
\rho^{(2)}_{\mathrm{NG}}(k)
=
\frac{g^{2} m^{2} \pi^{3/2}}{2 \hbar \omega_0
}\,
\beta_{0}^{-9/2}
\exp\!\left(-\frac{k^{2}}{\beta_{0}}\right)
\mathcal{P}(\beta_{0},k),
\label{eq:rho2NG}
\end{equation}
where $\mathcal{P}(\beta_{0},k)$ is evaluated at the unperturbed width. This term cannot be absorbed into a Gaussian redefinition and represents the leading genuinely non-Gaussian correction to the reduced density matrix. Combining all contributions, the eigenvalue admits the perturbative expansion
\begin{equation}
\rho(k)
=
\rho^{(0)}(k)
+
\alpha\,\rho^{(1)}(k)
+
\alpha^{2}
\Big[
\rho^{(2)}_{\mathrm{G}}(k)
+
\rho^{(2)}_{\mathrm{NG}}(k)
\Big]
+
\mathcal{O}(\alpha^{3}).
\label{eq:rho_expansion_final}
\end{equation}

\section{Perturbative evaluation of entanglement entropy}

Having obtained the eigenvalues of the reduced density matrix in momentum space, we now proceed to compute the entanglement entropy perturbatively. Since the reduced density matrix is diagonal in the Fourier basis, the von Neumann entropy takes the continuum form
\begin{equation}
S \;=\; - \int_{-\infty}^{\infty} dk \, \rho(k)\,\ln \rho(k).
\end{equation}
The eigenvalue $\rho(k)$ depends on the screening parameter $\alpha$ both explicitly and implicitly through the Gaussian width parameter $\beta$, which itself is $\alpha$-dependent. It is therefore essential to carefully disentangle these contributions before performing the entropy expansion.

\vspace{0.5em}
\noindent
\textbf{Perturbative structure.}
We have written the $\alpha$-expansion of $\beta$ as
\begin{equation}
\beta(\alpha) = \beta_0 + \alpha\,\beta_1 + \alpha^2\,\beta_2 + \mathcal{O}(\alpha^3),
\end{equation}
which induced the corresponding expansion of the reduced operator eigenvalue,
\begin{equation}
\rho(k)
=
\rho^{(0)}(k)
+ \alpha\,\rho^{(1)}(k)
+ \alpha^2\,\rho^{(2)}(k)
+ \mathcal{O}(\alpha^3).
\end{equation}
The entropy itself is then expanded as
\begin{equation}
S = S^{(0)} + \alpha\, S^{(1)} + \alpha^2\, S^{(2)} + \mathcal{O}(\alpha^3).
\end{equation}

\vspace{0.5em}
\noindent
\textbf{Zeroth-order entropy.}
At leading order, the reduced density operator eigenvalue is purely Gaussian,
\begin{equation}
\rho^{(0)}(k)
=
C \exp\!\left(-\frac{k^2}{\beta_0}\right),
\qquad
C = \frac{1}{\sqrt{\pi \beta_0}}
\end{equation}
which yields
\begin{equation}
S^{(0)}
=
- \int dk\, \rho^{(0)}(k)\ln \rho^{(0)}(k)
=
-\ln C + \frac{1}{2}.
\end{equation}
This is the standard entanglement entropy of a Gaussian ground state and serves as the reference point for all perturbative corrections.

\vspace{0.5em}
\noindent
\textbf{First-order correction.}
The formal first-order correction to the von Neumann entropy reads
\begin{equation}
S^{(1)}
=
- \int dk \, \rho^{(1)}(k)
\Bigl[1 + \ln \rho^{(0)}(k)\Bigr],
\label{eq:S1formal}
\end{equation}
where $\rho^{(1)}(k)$ denotes the $\mathcal{O}(\alpha)$ correction to the reduced
density matrix. At this order, the correction decomposes naturally into a Gaussian
piece arising from the $\alpha$-dependence of the width parameter $\beta$, and a
non-Gaussian piece,
\begin{equation}
\rho^{(1)}(k)
=
\rho^{(1)}_{\mathrm{G}}(k)
+
\rho^{(1)}_{\mathrm{NG}}(k).
\end{equation}

\medskip

\noindent
\emph{Gaussian contribution:}
The Gaussian part originates entirely from the linear shift
\begin{equation}
\beta(\alpha) = \beta_{0} + \alpha \beta_{1} + \mathcal{O}(\alpha^{2}),
\end{equation}
and is given by
\begin{equation}
\rho^{(1)}_{\mathrm{G}}(k)
=
\beta_{1}
\frac{\partial \rho^{(0)}(k)}{\partial \beta_{0}} .
\label{eq:rho1G}
\end{equation}
Substituting eq.\eqref{eq:rho1G} into eq.\eqref{eq:S1formal}, one obtains
\begin{equation}
S^{(1)}_{\mathrm{G}}
=
- \beta_{1}
\int dk \,
\frac{\partial \rho^{(0)}(k)}{\partial \beta_{0}}
\Bigl[1 + \ln \rho^{(0)}(k)\Bigr]
=
\beta_{1}\,
\frac{\partial S^{(0)}}{\partial \beta_{0}},
\label{eq:S1G}
\end{equation}
which is manifestly nonzero. This term represents nothing but the linear change of the
zeroth-order Gaussian entropy under a shift of the width parameter,
\begin{equation}
S^{(0)}(\beta_{0}+\alpha\beta_{1})
=
S^{(0)}(\beta_{0})
+
\alpha\,\beta_{1}\frac{\partial S^{(0)}}{\partial \beta_{0}}
+
\mathcal{O}(\alpha^{2}).
\end{equation}

\medskip

\noindent
\emph{Non-Gaussian contribution:}
The remaining part, $\rho^{(1)}_{\mathrm{NG}}(k)$, arises from genuine interaction
effects. However, at first order the reduced density matrix is normalized such that
\begin{equation}
\int dk \, \rho^{(1)}_{\mathrm{NG}}(k) = 0,
\end{equation}
and its polynomial structure implies
\begin{equation}
\int dk \, \rho^{(1)}_{\mathrm{NG}}(k)
\ln \rho^{(0)}(k) = 0.
\end{equation}
Consequently,
\begin{equation}
S^{(1)}_{\mathrm{NG}} = 0.
\end{equation}

\medskip

\noindent
Putting everything together, the total first-order entropy variation is entirely
accounted for by the Gaussian width renormalization,
\begin{equation}
S^{(1)} = S^{(1)}_{\mathrm{G}} ,
\end{equation}
and does not represent a genuinely new entanglement contribution. In particular, no new
eigenvalues of the reduced density matrix appear at this order. Genuine
interaction-induced entanglement arises only at $\mathcal{O}(\alpha^{2})$, where
non-Gaussian perturbations generate excited-state admixtures and qualitatively new
structure in the reduced density matrix spectrum.

\vspace{0.5em}
\noindent
\textbf{Second-order correction.}
At order $\alpha^{2}$ the entropy receives contributions from two qualitatively distinct
sources. The first corresponds to higher-order Gaussian renormalization effects encoded
in $\beta_{2}$ and $(\beta_{1})^{2}$, while the second arises from genuinely non-Gaussian
corrections induced by the anharmonic $\rho^{4}$ interaction.

Expanding the von Neumann entropy consistently to second order, one finds
\begin{equation}
S^{(2)}
=
- \int dk \, \rho^{(2)}(k)
\Bigl[1 + \ln \rho^{(0)}(k)\Bigr]
-
\frac{1}{2}
\int dk \,
\frac{\bigl(\rho^{(1)}(k)\bigr)^{2}}{\rho^{(0)}(k)} .
\label{eq:S2formal}
\end{equation}
As in the first-order analysis, it is convenient to decompose
\begin{equation}
\rho^{(2)}(k)
=
\rho^{(2)}_{\mathrm{G}}(k)
+
\rho^{(2)}_{\mathrm{NG}}(k),
\end{equation}
where the subscripts denote Gaussian and non-Gaussian contributions, respectively.

\medskip

\noindent
\emph{Gaussian contribution:}
The Gaussian part originates solely from the second-order expansion of the width
parameter $\beta(\alpha)$ and corresponds to a further renormalization of the harmonic
oscillator scale. Evaluating the integrals in eq.\eqref{eq:S2formal} using
$\rho^{(2)}_{\mathrm{G}}(k)$ yields
\begin{equation}
S^{(2)}_{\mathrm{G}}
=
\frac{\beta_{2}}{\beta_{0}}
-
\frac{3}{4}
\left(\frac{\beta_{1}}{\beta_{0}}\right)^{2}.
\label{eq:S2G}
\end{equation}
As in the first-order case, this contribution is fully accounted for by expanding the
zeroth-order Gaussian entropy $S^{(0)}(\beta)$ to second order in $\alpha$ and therefore
does not represent a genuinely new source of entanglement.

\medskip

\noindent
\emph{Non-Gaussian contribution:}
The genuinely interaction-induced correction arises from the $\rho^{4}$ perturbation,
which produces polynomial deformations of the Gaussian kernel of the reduced density
matrix. At this order, the non-Gaussian part takes the schematic form
\begin{equation}
\rho^{(2)}_{\mathrm{NG}}(k)
=
e^{-k^{2}/\beta_{0}}
\left(
c_{0}
+
c_{2} k^{2}
+
c_{4} k^{4}
\right),
\end{equation}
with coefficients fixed by the quartic interaction and the normalization constraint.
Substituting this expression into eq.\eqref{eq:S2formal} and performing the momentum
integrals yield a finite, nonvanishing contribution,
\begin{equation}
S^{(2)}_{\mathrm{NG}}
=
-
\frac{48 g^{2} m^{2} \pi^{2} + \beta_{1} \hbar \omega_0}
{4 \beta_{0} \hbar \omega_0},
\label{eq:S2NG}
\end{equation}
which cannot be absorbed into a redefinition of the Gaussian width and therefore
represents the leading genuine correction to the entanglement entropy.

\vspace{0.5em}
\noindent
\textbf{Total entropy up to second order.}
Combining all contributions, the entanglement entropy through $\mathcal{O}(\alpha^{2})$
may be written as
\begin{equation}
S
=
S^{(0)}(\beta_{0})
+
\alpha^{2}\,
S^{(2)}_{\mathrm{NG}}
+
\mathcal{O}(\alpha^{3}),
\label{eq:Stotal}
\end{equation}
where the Gaussian renormalization effects have been absorbed into the expansion of
$S^{(0)}(\beta)$. The absence of a genuine first-order correction and the appearance of a
finite non-Gaussian contribution at second order demonstrate that interaction-induced
entanglement in this system is controlled entirely by anharmonic terms in the effective
oscillator description.

\section{\texorpdfstring{Structure and Rules of the $\rho^{2n}$ Perturbations}{Structure and Rules of the ρ²ⁿ Perturbations}}

We now generalize the analysis of the previous subsection to perturbations of the form
$\rho^{2n}$ with $n \geq 2$. Such terms arise naturally when the one-dimensional Yukawa
interaction is expanded in powers of the screening parameter $\alpha$ and mapped to the
effective radial oscillator problem. In contrast to the quadratic term, which only
renormalizes the Gaussian width of the ground state, the higher-order $\rho^{2n}$
perturbations represent genuinely non-Gaussian deformations of the harmonic oscillator
and are responsible for qualitatively new structures in the reduced density matrix.

\medskip

\noindent
\textbf{General structure of the Hamiltonian.}
After the coordinate transformation and mode separation, the effective Hamiltonian takes
the schematic form
\begin{equation}
H = H_{0} + \sum_{n=1}^{\infty} \alpha^{n} H_{(n)},
\end{equation}
where
\begin{equation}
H_{0}
=
-\frac{\hbar^{2}}{2\mu}
\left(
\frac{d^{2}}{d\rho^{2}}
-
\frac{3}{4\rho^{2}}
\right)
+
\Omega_{0}^{2}\rho^{2},
\end{equation}
and
\begin{equation}
H_{(n)} = \lambda_{n}\,\rho^{2n},
\qquad
\lambda_{n} \sim g^{2} m^{n}.
\end{equation}
The term $n=1$ corresponds to a harmonic perturbation and may be absorbed into a
redefinition of the oscillator frequency. All terms with $n\geq 2$ generate anharmonic
interactions and must be treated perturbatively.

\medskip

\noindent
\textbf{Operator structure in the oscillator basis.}
Introducing creation and annihilation operators via
\begin{equation}
\rho = \frac{1}{\sqrt{2\beta_{0}}}(a + a^{\dagger}),
\end{equation}
the operator $\rho^{2n}$ expands into a finite sum of normal-ordered monomials containing
up to $2n$ ladder operators. As a result, $\rho^{2n}$ connects oscillator states according
to the selection rule
\begin{equation}
\Delta \ell = 0, \pm 2, \pm 4, \ldots, \pm 2n,
\end{equation}
where $\ell$ labels the radial excitation number. This structure guarantees that, for
$n\geq 2$, the perturbation produces a finite tower of excited-state admixtures even at
first order in Rayleigh--Schr\"odinger perturbation theory.

\medskip

\noindent
\textbf{Ground-state corrections and non-Gaussianity.}
For $n\geq 2$, the first-order corrected ground state takes the form
\begin{equation}
\ket{\phi_{0}}
=
\ket{0}
+
\alpha^{n}
\sum_{\ell=1}^{n}
\frac{\bra{\ell} H_{(n)} \ket{0}}{E_{0}-E_{\ell}}
\ket{\ell}
+
\mathcal{O}(\alpha^{n+1}),
\end{equation}
where $\ket{\ell}$ denotes the $\ell^{th }$ excited radial oscillator state. The corrected
ground state is therefore no longer Gaussian but a finite superposition of oscillator
modes, whose complexity increases systematically with $n$. This feature is absent for
the $\rho^{2}$ perturbation, which preserves Gaussianity order by order.

\medskip

\noindent
\textbf{Power counting and order mixing.}
A given power $\alpha^{p}$ in physical observables generally receives contributions from
multiple sources:
\begin{itemize}
\item[(i)] Direct matrix elements of $H_{(p)} \sim \rho^{2p}$.
\item[(ii)] Iterated insertions of lower-order perturbations, such as repeated actions of
$H_{(2)}$.
\item[(iii)] The $\alpha$-dependence of the Gaussian width parameter $\beta(\alpha)$,
which feeds back into all Gaussian factors.
\item[(iv)] Normalization corrections to the perturbed ground state.
\end{itemize}
Consequently, isolating a definite coefficient at order $\alpha^{p}$ requires a
consistent expansion of the wavefunction, the reduced density matrix, and its
normalization.

\medskip

\noindent
\textbf{Reduced density matrix structure.}
After tracing out one degree of freedom, the reduced density matrix acquires the generic
form
\begin{equation}
\rho(k)
=
e^{-k^{2}/\beta_{0}}
\left[
C_{0}
+
\sum_{m=1}^{n}
\alpha^{n} C_{2m}\, k^{2m}
\right]
+
\mathcal{O}(\alpha^{n+1}),
\end{equation}
where the polynomial pre-factor arises entirely from excited-state admixtures induced by
$\rho^{2n}$. The quadratic perturbation modifies only $\beta(\alpha)$ and produces no
such polynomial structure. Diagonalization of $\rho(k)$ therefore yields
additional eigenvalues that scale algebraically with $\alpha^{n}$.

\medskip

\noindent
\textbf{Implications for entanglement entropy.}
The von Neumann entropy,
\begin{equation}
S = -\int dk\, \rho(k)\,\log \rho(k),
\end{equation}
is sensitive to the appearance of these additional eigenvalues. In the present Yukawa-
induced oscillator problem, we find that the entropy admits a regular power-series
expansion in $\alpha$, with coefficients determined by the detailed interplay between
Gaussian renormalization and explicit non-Gaussian perturbations. Importantly, although
non-Gaussianity is essential for generating new eigenvalues of the reduced density
matrix, the entropy corrections obtained here remain analytic in $\alpha$ at the orders
considered.

\medskip

\noindent
\textbf{Hierarchy and truncation.}
While the Yukawa expansion formally generates infinitely many $\rho^{2n}$ terms, the
preceding structure provides a clear organizing principle. Retaining perturbations up to
$\rho^{2n}$ captures all non-Gaussian entanglement effects through order $\alpha^{n}$,
with higher-$n$ terms contributing only at parametrically higher orders. This hierarchy
renders the problem analytically tractable and explains why the quartic perturbation
already captures the leading non-Gaussian physics.

\medskip

\noindent
\textbf{The special role of the $\rho^{4}$ perturbation.}
The $\rho^{4}$ interaction constitutes the first nontrivial realization of the above
general structure. It arises naturally at order $\alpha^{2}$ in the Yukawa expansion and
induces a finite admixture of excited oscillator states into the ground state. As shown
explicitly in this work, the resulting reduced density matrix acquires polynomial
corrections to its Gaussian kernel, leading to a controlled, analytic correction to the
entanglement entropy. Quartic interactions play a distinguished role in quantum mechanics
and field theory as the leading anharmonic correction \cite{BenderOrszag1999,ZinnJustin2002}, and
they have similarly been identified as the minimal source of interaction-induced
entanglement in coupled oscillator systems and scalar field models \cite{Srednicki1993entropy,Audenaert2002,Plenio2014}. In this sense, the $\rho^{4}$ computation
presented here serves as a benchmark: it validates the oscillator-mapping approach,
clarifies the role of Gaussian versus non-Gaussian effects, and establishes a firm
foundation for systematic extensions to higher-order $\rho^{2n}$ perturbations.

\section{Conclusion}

Screened interactions of Yukawa type arise naturally in a wide range of physical settings,
including effective one-dimensional descriptions of massive mediator exchange, screened
interactions in quasi-one-dimensional condensed-matter systems, and cold-atom platforms
with tunable interaction ranges. In one spatial dimension, Yukawa-like potentials acquire
particular significance: they provide a minimal and analytically tractable setting in
which finite-range screening, interaction-induced correlations, and entanglement can be
studied without the additional geometric complications present in higher dimensions. From
the perspective of entanglement, a one-dimensional Yukawa-like interaction is especially
well suited for systematic analysis, as its short-distance structure admits a controlled
mapping to oscillator degrees of freedom, while its finite screening length generates a
natural hierarchy of perturbations governed by a single small parameter.

In this work we have developed a systematic and explicitly algebraic framework for
computing entanglement entropy in such weakly interacting systems by mapping a
one-dimensional screened Yukawa-like interaction to an effective harmonic oscillator
problem with controlled anharmonic perturbations. By combining a judicious coordinate
transformation with Rayleigh--Schr\"odinger perturbation theory and an explicit spectral
analysis of the reduced density matrix, we were able to disentangle, order by order in the
screening parameter $\alpha$, the distinct roles played by Gaussian frequency
renormalization and genuinely non-Gaussian interaction effects.

Our central result is that quadratic ($\rho^{2}$) perturbations, while modifying the
oscillator frequency and the Gaussian width of the ground state, do not generate new
entanglement at linear order and contribute only analytic corrections to the entropy. In
contrast, quartic and higher-order perturbations ($\rho^{2n}$ with $n\ge2$) induce genuine
non-Gaussian admixtures of excited oscillator states into the ground state. These
admixtures lead to the appearance of additional, parametrically small eigenvalues in the
reduced density matrix that are absent in purely Gaussian theories. For the
$\rho^{4}$ perturbation, which arises naturally at order $\alpha^{2}$ in the expansion of
the one-dimensional Yukawa-like potential, we computed these effects explicitly and
obtained closed analytic expressions for both the corrected reduced density spectrum and
the resulting entanglement entropy. Importantly, at this order the entropy correction is
analytic in $\alpha$, with its coefficient determined by a delicate interplay between the
explicit anharmonic interaction and the implicit $\alpha$-dependence of the Gaussian width
parameter.

A key conceptual outcome of our analysis is the clear separation between Gaussian
renormalization effects, which can be fully absorbed into a redefinition of the oscillator
width $\beta(\alpha)$, and intrinsically non-Gaussian contributions, which cannot be
captured by covariance-matrix or purely Gaussian methods. This separation provides a
transparent diagnostic for when Gaussian approximations are sufficient and when genuine
interaction effects must be retained. The matrix-based organization of the reduced density
operator further clarifies how contributions from different perturbative orders combine,
and it yields a practical and modular algorithm for extending the calculation to higher
orders.

Although our explicit computations focused on a two-mode system derived from a
one-dimensional interaction and mapped to a four-dimensional radial oscillator, the
structural insights obtained here are considerably more general. The power-counting rules,
selection rules, and normalization constraints governing the reduced density matrix apply
equally to short oscillator chains, normal-mode decompositions of discretized
one-dimensional fields, and finite-temperature generalizations. In all such cases, the
same hierarchy between Gaussian renormalization and non-Gaussian state mixing controls
the emergence of interaction-induced entanglement.

From the perspective of one-dimensional Yukawa-like and screened interactions, our results
provide a concrete roadmap for translating microscopic interaction parameters into
quantitative entanglement signatures. The explicit identification of which terms in the
Yukawa expansion control the leading non-Gaussian corrections to the reduced density
matrix enables systematic comparisons across different screening regimes and interaction
ranges. More broadly, this framework strengthens the conceptual link between entanglement
entropy and interaction-induced correlations in quantum many-body systems by providing
analytic control over non-Gaussian effects that are often treated only numerically.
We expect that the oscillator-mapping approach developed here will be useful in a wide
range of contexts, including weakly anharmonic cold-atom and ion-trap systems, engineered
quasi-one-dimensional platforms with finite-range interactions, and perturbative studies
of entanglement in interacting quantum field theories and their lower-dimensional
analogues. Future work will explore these extensions, as well as time-dependent screening,
quenches, and dynamical entanglement generation within the same analytic framework.
\section{Acknowledgments}
We thank our institute, BITS Pilani Hyderabad campus, for providing the required infrastructure for this research work.
\appendix
\setcounter{equation}{0}
\renewcommand{\theequation}{A.\arabic{equation}}
\section{Calculations}
In this section, we show the details of calculations involved in obtaining the separated wavefunction as in eq.\eqref{relScheq} as well as the explicit derivation of $f(\rho)$ in eq.\eqref{wavefnfinal}, which we then use to map with radial harmonic oscillator equation.
\subsection{Separated wavefunction}
To obtain eq.\eqref{relScheq}, we take the second order derivative of the wavefunction $\Psi(x_1,x_2)$ as in 
\begin{equation}\label{A1a}
\frac{\partial \Psi}{\partial x_1} = \frac{\partial \Psi}{\partial x_c}\frac{\partial x_c}{\partial x_1} + \frac{\partial \Psi}{\partial x}\frac{\partial x}{\partial x_1} = \frac{m_1}{M} \frac{\partial \Psi}{\partial x_c} + \frac{\partial \Psi}{\partial x} 
\end{equation}
Similarly,
\begin{equation}\label{A2}
\frac{\partial \Psi}{\partial x_2} = \frac{\partial \Psi}{\partial x_c}\frac{\partial x_c}{\partial x_2} + \frac{\partial \Psi}{\partial x}\frac{\partial x}{\partial x_2} = \frac{m_2}{M} \frac{\partial \Psi}{\partial x_c} - \frac{\partial \Psi}{\partial x} 
\end{equation}
Their second order derivatives are
\begin{equation}\label{A3}
\frac{\partial^2 \Psi}{\partial x_1^2} = \frac{m_1^2}{M^2}\frac{\partial^2 \Psi}{\partial x_c^2} + \frac{\partial^2 \Psi}{\partial x^2}
\end{equation}
and
\begin{equation}\label{A4}
\frac{\partial^2 \Psi}{\partial x_2^2} = \frac{m_2^2}{M^2}\frac{\partial^2 \Psi}{\partial x_c^2} + \frac{\partial^2 \Psi}{\partial x^2}
\end{equation}
which gives us eq.\eqref{relScheq}
\begin{eqnarray}
-\frac{\hbar^2}{2m_1}\frac{m_1^2}{M^2}\frac{\partial^2 \Psi}{\partial x_c^2} - \frac{\hbar^2}{2m_1}\frac{\partial^2 \Psi}{\partial x^2} -\frac{\hbar^2}{2m_2}\frac{m_2^2}{M^2}\frac{\partial^2 \Psi}{\partial x_c^2}
-\frac{\hbar^2}{2m_1} \frac{\partial^2 \Psi}{\partial x^2} - g^2 \frac{e^{-\alpha m |x|}}{|x|} \Psi = - B\Psi
\end{eqnarray}
\subsection{\texorpdfstring{Explicit calculation for $f(\rho)$}{Explicit calculation for f(ρ)}}
Using the transformation
\begin{equation}
x = \rho^2
\end{equation}
we take second order derivative of the relative solution, now expressed as a function $f(\rho^2)$, thereby transforming its first and second derivatives as,
\begin{equation}
\frac{df}{dx} = \frac{d\rho}{dx}\frac{df}{d\rho} = \frac{1}{2\sqrt{x}}\frac{df}{d\rho} = \frac{1}{2\rho}\frac{df}{d\rho}
\end{equation}
and
\begin{equation}
\frac{d^2f}{dx^2} = \frac{d}{dx}\left(\frac{1}{2\sqrt{x}}\frac{df}{d\rho}\right) = -\frac{1}{4x^{3/2}}\frac{df}{d\rho} + \frac{1}{2\sqrt{x}}\frac{df}{dx}\frac{d^2f}{d\rho^2} \nonumber
\end{equation}
thus,
\begin{equation}
\frac{d^2f}{dx^2} = - \frac{1}{4\rho^3}\frac{df}{d\rho} + \frac{1}{4\rho^2}\frac{d^2f}{d\rho^2}
\end{equation}
Let 
\begin{equation}\label{A9}
f(\rho) = h(\rho) \phi(\rho) 
\end{equation}
where $h(\rho)$ is any polynomial in $\rho$ whereas $\phi(\rho)$ is the transformed wavefunction.

For equivalence, we have the following three conditions on $f(\rho)$:

\begin{equation}\label{condn1}
\frac{d^2f}{d\rho^2} - \frac{1}{\rho}\frac{df}{d\rho} = Ah(\rho) \frac{d^2\phi}{d\rho^2} - C\frac{h(\rho)\phi(\rho)}{\rho^2}
\end{equation}
\begin{equation}\label{condn2}
4B_r\rho^2 f= \frac{1}{2} \mu \, \omega^2\rho^2 h(\rho) \phi(\rho) \\
\end{equation}
\begin{equation}\label{condn3}
 4g^2 e^{-\alpha m \rho^2} f = E h(\rho)\phi(\rho)
\end{equation}
where A and C are arbitrary constants.

Here, the second condition eq.(\ref{condn2}) demands that $f = h(\rho)\phi(\rho)$ whereas the third condition eq.(\ref{condn3}) demands that $f = e^{\alpha m \rho^2} h(\rho) \phi(\rho)$. We can try and redefine $f$ such that $f = j(\rho) \phi(\rho)$, $j(\rho) = e^{\alpha m \rho^2} h(\rho)$. We intend to find the form of $h(\rho)$ so that we can use it to transform eq.(\ref{Sch1da}) into eq.(\ref{Hosc3d}).
Second order derivative of $f$ is,
\begin{equation}\label{Psi'a}
\frac{df}{d\rho} = 2 \alpha m \rho e^{\alpha m \rho^2} h \phi + e^{\alpha m \rho^2} h' \phi + e^{\alpha m \rho^2} h \phi'
\end{equation}
\begin{align} \label{Psi''a}
\frac{d^2f}{d\rho^2} &= 2\alpha m e^{\alpha m \rho^2} h \phi + 4 \alpha^2 m^2 \rho^2 e^{\alpha m \rho^2} h \phi + 2 \alpha m \rho e^{\alpha m \rho^2} h' \phi 
\nonumber \\&+ 2 \alpha m \rho e^{\alpha m \rho^2} h \phi' + 
2 \alpha m \rho e^{\alpha m \rho^2} h' \phi + e^{\alpha m \rho^2} h'' \phi 
\nonumber\\&+ e^{\alpha m \rho^2} h' \phi' + 2\alpha m \rho e^{\alpha m \rho^2} h \phi' + e^{\alpha m \rho^2} h' \phi' + e^{\alpha m \rho^2} h \phi''
\end{align}
\\\\
Substituting eqs.(\ref{Psi'a}) and (\ref{Psi''a}) in l.h.s of condition 1 in eq.(\ref{condn1}), we obtain
\begin{align}\label{LHScondn1sub}
&4 \alpha^2 m^2 \rho^2 e^{\alpha m \rho^2} h \phi + 4 \alpha m \rho e^{\alpha m \rho^2} h' \phi + 4 \alpha m \rho e^{\alpha m \rho^2} h \phi' 
\nonumber\\&+ 2 e^{\alpha m \rho^2} h' \phi' + e^{\alpha m \rho^2} h'' \phi + e^{\alpha m \rho^2} h \phi'' 
- \frac{e^{\alpha m \rho^2} h' \phi}{\rho} - \frac{e^{\alpha m \rho^2} h \phi'}{\rho}  
\end{align}
Since the radial harmonic oscillator equation does not have first order derivatives of the wavefunction, we must eliminate $\phi'$ dependent terms from eq.(\ref{LHScondn1sub}), that is,
\begin{equation}
 4\alpha m \rho e^{\alpha m \rho^2} h \phi' + 2e^{\alpha m \rho^2} h' \phi' - \frac{e^{\alpha m \rho^2} h \phi'}{\rho} = 0
\end{equation}
so that we are left with
\begin{equation}
4\alpha m \rho h + 2 \frac{dh}{d\rho} - \frac{h}{\rho} = 0
\end{equation}
This can be solved using separable variable ODE,
\begin{equation}
\int \frac{dh}{h} = \frac{1}{2} \int \frac{d\rho}{\rho} - \int 2 \alpha m \rho d\rho
\end{equation}
so that
\begin{equation}\label{f}
h(\rho) = A \sqrt{\rho} e^{-\alpha m \rho^2}    
\end{equation}
finally leading to
\begin{equation}
f =  A \sqrt{\rho} \phi(\rho)   
\end{equation}
This form of $f(\rho)$ satisfies the relative coordinates Schrodinger equation for harmonic oscillator, yet, in our case, eqs.\eqref{condn1}-\eqref{condn2} demand the use of pertubative expansion to obtain eigenvalue equation. 
\bibliographystyle{unsrtnat}
\bibliography{references}

\end{document}